\documentstyle[prl,floats,multicol,amssymb,amsgen,
amsfonts,amsbsy,aps]{revtex}

\tighten
\begin{document}
\draft
\title{Telecloning and multiuser quantum channels for continuous variables}
\author{P.\ van Loock and Samuel L.\ Braunstein}
\address{Informatics, Bangor University, Bangor LL57 1UT, UK}
\maketitle

\begin{abstract}
We propose entangled $(M+1)$-mode quantum states as a 
multiuser quantum channel for continuous-variable communication. 
Arbitrary quantum states can be sent via this channel {\it simultaneously} 
to $M$ remote and separated locations with equal minimum excess noise 
in each output mode.
For a set of coherent-state inputs, the channel realizes optimum 
symmetric $1\rightarrow M$ cloning at a distance (``telecloning'').
It also provides the optimal cloning of coherent states without the need
of amplifying the state of interest. 
The generation  of the multiuser quantum channel requires no more than
two $10\log_{10}[(\sqrt{M}-1)/(\sqrt{M}+1)]$ dB squeezed states and
$M$ beam splitters. 
\end{abstract}
\vspace{3ex}

\begin{multicols}{2}

Quantum information encoded in non-orthogonal quantum states can be
perfectly transferred between two distant locations that are linked
by a maximally entangled state and a classical communication
channel. This quantum teleportation \cite{Benn1}
is a prime example of quantum information processing
\cite{Nielsen} where otherwise impossible cryptographic,
computational, and communication tasks can be performed through
the presence of shared entanglement. In principle, perfect teleportation 
with unit fidelity from a sender to a {\it single} receiver is possible 
in accordance with quantum mechanics. What about conveying quantum
information via a ``multiuser quantum channel'' (MQC) simultaneously to
several receivers? Of course, the
no-cloning theorem \cite{Woott} of quantum theory forbids {\it perfect} 
cloning (or copying) of unknown non-orthogonal quantum states and does so
also over a distance. This prevents the MQC from being able
to produce {\it exact} clones of the sender's input state at all
receiving stations. The MQC, however, can provide
each receiver with at least a part of the input quantum information and
distribute {\it approximate} clones with non-unit fidelity \cite{Buzek}.
This cloning at a distance or ``telecloning'' may be 
seen as the ``natural generalization of teleportation to the many-recipient 
case'' \cite{Murao}.

Whereas the original teleportation proposal \cite{Benn1} had 
been extended to infinite-dimensional Hilbert 
spaces \cite{Vaid,Sam1} followed by a successful demonstration of
continuous-variable teleportation \cite{Furu}, most results on 
quantum cloning refer to finite-dimensional systems, in particular, qubits
\cite{Buzek,Gisin,Bruss1,Bruss2,Werner}. A cloning experiment has been proposed
for single-photon qubits \cite{Simon}, and two other optical qubit cloning
experiments have been realized
\cite{Martini,Wan}. For qubits, telecloning has also
been studied theoretically, first with one input sent to two
receivers \cite{Bruss1}, and more generally with one input 
\cite{Murao} and $N$ identical inputs \cite{Duer} distributed among
$M$ receivers. The telecloning scenario with one input copy and
$M$ receivers has then been extended to $d$-level systems \cite{Murao2}. 

The first investigations on continuous-variable cloning led to
cloning transformations for certain sets of input states, namely
coherent (or certain squeezed) states, enabling optimum ``local''
$1\rightarrow 2$ cloning (one state mapped to two approximate copies)
with fidelity
$F_{{\rm clon},1,2}^{{\rm coh\,st},\infty}=2/3$ \cite{Cerf1}.
Subsequently, fidelity boundaries of local Gaussian $N\rightarrow M$ cloners
were derived, $F_{{\rm clon},N,M}^{{\rm coh\,st},\infty}\leq MN/(MN+M-N)$
\cite{Cerf2}. In Ref.~\onlinecite{Sam2}, it has been shown that for any
Hilbert space dimension, the optimal symmetric universal local cloner, 
that clones all possible input states equally well, 
can be constructed from a single family of quantum circuits. In the
continuous limit, this universal cloner simply reduces to a classical
probability distributor attaining $F_{{\rm clon},N,M}^{{\rm univ},\infty}=N/M$
\cite{Sam2}, consistent with the limit of Werner's $d$-dimensional result,
$F_{{\rm clon},N,M}^{{\rm univ},d}=[N(d-1)+M(N+1)]/M(N+d)$ \cite{Werner}.
In fact, the local $N\rightarrow M$ coherent-state cloner
turns out to be a simple classical amplitude distributor which
can be built from a phase-insensitive amplifier and beam splitters
\cite{local} (see also \cite{dariano}).
We will here propose an MQC capable of optimum symmetric 
$1\rightarrow M$ telecloning of coherent states. More generally, 
this MQC,
produceable with squeezed light and linear optics, 
transfers arbitrary quantum states from a sender to $M$ receivers with 
equal minimum excess noise in each output state. 
Further, it forms a cloning circuit with no need to amplify the 
input.

Clearly a telecloner needs entanglement as soon its fidelity
is greater than the maximum fidelity attainable by classical teleportation
$F_{\rm class}$. In fact, for {\it universal} $1\rightarrow M$ qubit cloning
we have $F_{{\rm clon},1,M}^{{\rm univ},2}>F_{\rm class}=2/3$
\cite{Gisin,Bruss2,Werner,Popes}, for $1\rightarrow M$ cloning of 
{\it coherent states}
we have $F_{{\rm clon},1,M}^{{\rm coh\,st},\infty}>F_{\rm class}=1/2$ 
\cite{Cerf2,Sam3}.  
Therefore, optimum telecloning cannot be achieved by simply measuring
the input state and sending copies of the classical
result to all receivers.
On the other hand, in the limit $M\to\infty$, both
$F_{{\rm clon},1,M}^{{\rm univ},2}\to F_{\rm class}=2/3$ and
$F_{{\rm clon},1,M}^{{\rm coh\,st},\infty}\to F_{\rm class}=1/2$
which implies that {\it no} entanglement is needed for
infinitely many copies.

The most wasteful scheme
would be a protocol in which the sender locally creates $M$ optimum
clones and perfectly teleports one clone to each receiver using $M$
maximally entangled two-party states \cite{Murao,Murao2}. 
In fact, a much more economical 
strategy is that all participants share a particular multipartite
entangled state as a quantum channel. This MQC 
may contain maximum bipartite
entanglement ($\log_{2}d$ ebits) between the sender and all receivers
as does the $d$-level telecloning state of Murao {\it et al.} 
\cite{Murao,Murao2} (even in the limit $d \to\infty$).
The entanglement of the qubit state ($d=2$), $\log_{2}2=1$ ebit
\cite{Murao}, however, is larger than we expect 
from the most frugal scheme (it should become vanishingly small
as $M\to\infty$).
In a continuous-variable scenario based on the quadratures 
of single electromagnetic modes, multipartite entangled states can be
generated using squeezers and beam splitters \cite{PvL1,PvL2}, and
any maximum bipartite entanglement involved would require two sources 
of infinite squeezing.

We know now that, as opposed to the local continuous-variable cloners
which do not need any entanglement as a basic ingredient \cite{Sam2,local},
a telecloner must have entanglement. The following pure
$(M+1)$-mode Wigner function describes an appropriate candidate
for an MQC, since it enables optimum
$1\rightarrow M$ telecloning of coherent states:
\end{multicols}
\noindent\rule{5cm}{.6pt}
\begin{eqnarray}\label{MQC}
W_{\rm MQC}({\bf x},{\bf p})
=\left(\frac{2}{\pi}\right)^{M+1}
\exp\Bigg\{
&-&2e^{-2(s+r_1)}\left(\sin \theta_0\,x_1+
\frac{\cos \theta_0}{\sqrt{M}}\sum_{i=2}^{M+1} x_i \right)^2
-2e^{+2(s+r_1)}\left(\sin \theta_0\,p_1+
\frac{\cos \theta_0}{\sqrt{M}}\sum_{i=2}^{M+1} p_i \right)^2\nonumber\\
&-&2e^{-2(s-r_2)}\left(\cos \theta_0\,x_1-
\frac{\sin \theta_0}{\sqrt{M}}\sum_{i=2}^{M+1} x_i \right)^2
-2e^{+2(s-r_2)}\left(\cos \theta_0\,p_1-
\frac{\sin \theta_0}{\sqrt{M}}\sum_{i=2}^{M+1} p_i \right)^2\nonumber\\
&-&\frac{1}{M}\sum_{i,j=2}^{M+1}
\left[e^{-2s}(x_i-x_j)^2+e^{+2s}(p_i-p_j)^2\right]
\Bigg\} \;,
\end{eqnarray}
\hfill\rule{5cm}{.6pt}
\begin{multicols}{2}
\noindent where ${\bf x}=(x_1,x_2,...,x_{M+1})$, 
${\bf p}=(p_1,p_2,...,p_{M+1})$, and
\begin{eqnarray}\label{conditions}
\frac{1}{\sqrt{M+1}}&\leq& \sin \theta_0 \leq \sqrt{\frac{M}{M+1}} \;,\\
e^{-2r_1}&=&\frac{\sqrt{M} \sin \theta_0 - \cos \theta_0}
{\sqrt{M} \sin \theta_0 + \cos \theta_0}\;,\\
e^{-2r_2}&=&\frac{\sqrt{M} \cos \theta_0 - \sin \theta_0}
{\sqrt{M} \cos \theta_0 + \sin \theta_0}\;.
\end{eqnarray}
We will explain the meaning of the different parameters in $W_{\rm MQC}$
later and first look at the potential telecloning protocol in which 
$W_{\rm MQC}$ is used. Let us assume $s=0$ and $\sin \theta_0=1/\sqrt{2}$. 
Mode 1 may be used as a ``port'' at the sending station and is combined at
a phase-free symmetric beam splitter with mode ``in'' which is in an 
{\it arbitrary} quantum state described by $W_{\rm in}$. The whole system 
after the beam splitter (we call the two modes emerging from the
beam splitter $\alpha_{\rm u}=x_{\rm u}+ip_{\rm u}$ and 
$\alpha_{\rm v}=x_{\rm v}+ip_{\rm v}$) can be written as
\begin{eqnarray}\label{afterBS} 
&W&(\alpha_{\rm u},\alpha_{\rm v},\alpha_2,...,\alpha_{M+1})=
\int dx_{\rm in}dp_{\rm in} W_{\rm in}(x_{\rm in},p_{\rm in})\nonumber\\ 
&\times&\, W_{\rm MQC}\left[\alpha_1=
\case{1}{\sqrt{2}}(\alpha_{\rm v}-\alpha_{\rm u}),
\alpha_2,...,\alpha_{M+1}\right]\nonumber\\
&\times&\, \delta\left[\case{1}{\sqrt{2}}(x_{\rm u}+x_{\rm v})
-x_{\rm in}\right]\,\delta\left[\case{1}{\sqrt{2}}
(p_{\rm u}+p_{\rm v})-p_{\rm in}\right] \;.
\end{eqnarray}
The ``Bell detection'', i.e., homodyne detections of 
$x_{\rm u}=\case{1}{\sqrt{2}}(x_{\rm in}-x_{\rm 1})$ and
$p_{\rm v}=\case{1}{\sqrt{2}}(p_{\rm in}+p_{\rm 1})$ \cite{Sam1}
can be described by the unnormalized reduced Wigner function
after integrating over $x_{\rm v}$ and $p_{\rm u}$:
\begin{eqnarray}\label{reduced}
&\propto& \int dx\, dp\, W_{\rm in}(x,p)\\
&\times& W_{\rm MQC}\left[x-\sqrt{2}x_{\rm u}+i(\sqrt{2}p_{\rm v}-p),
\alpha_2,...,\alpha_{M+1}\right].\nonumber
\end{eqnarray}
The $M$ distant and separated locations of modes 2 through $M+1$
{\it each} need to be provided now with the classical information of
the measurement results. Finally, when ``displacing'' all these modes
as $x_{\rm 2...M+1}\longrightarrow x_{\rm 2...M+1}+\sqrt{2}x_{\rm u}$ and
$p_{\rm 2...M+1}\longrightarrow p_{\rm 2...M+1}+\sqrt{2}p_{\rm v}$,
we obtain the ensemble description of the $M$-mode output
Wigner function after integrating out $x_{\rm u}$ and $p_{\rm v}$
for an ensemble of input states \cite{Sam1}
\end{multicols}
\noindent\rule{5cm}{.6pt}
\begin{eqnarray}\label{output}
W_{\rm out}(\alpha_2,...,\alpha_{M+1})
=\frac{2^M}{\pi^M(M-1)}
\exp\Bigg\{
&+&\frac{1}{M}\Bigg[\Bigg(\sum_{i=2}^{M+1} x_i \Bigg)^2-
\sum_{i,j=2}^{M+1} (x_i-x_j)^2+
\Bigg(\sum_{i=2}^{M+1} p_i \Bigg)^2-
\sum_{i,j=2}^{M+1} (p_i-p_j)^2 \Bigg]\Bigg\}\nonumber\\
\times\,\int dx\, dp\, W_{\rm in}(x,p)\,
\exp\Bigg\{
&-&\frac{1}{2}\frac{\sqrt{M}+1}{\sqrt{M}-1}\Bigg[\Bigg(x-
\frac{1}{\sqrt{M}}\sum_{i=2}^{M+1} x_i \Bigg)^2
+\Bigg(p-\frac{1}{\sqrt{M}}\sum_{i=2}^{M+1} p_i \Bigg)^2\Bigg]\\
&-&\frac{1}{2}\frac{\sqrt{M}-1}{\sqrt{M}+1}\Bigg[\Bigg(x+
\frac{1}{\sqrt{M}}\sum_{i=2}^{M+1} x_i \Bigg)^2+
\Bigg(p+\frac{1}{\sqrt{M}}\sum_{i=2}^{M+1} p_i \Bigg)^2\Bigg]
-(x^2+p^2)
\Bigg\}\;.\nonumber
\end{eqnarray}
%\hfill\rule{5cm}{.6pt}
\begin{multicols}{2}
\noindent This Wigner function is totally symmetric with respect
to all $M$ modes. We can therefore choose an arbitrary mode
and trace out (integrate out) the remaining $M-1$ modes which leaves 
us with the one-mode Wigner function of each individual clone 
\begin{eqnarray}\label{clone}
{\rm Tr}_{3...M+1}W_{\rm out}(\alpha_2,...,\alpha_{M+1})=
W_{\rm clon}(\alpha_2)\equiv W_{\rm clon}(\alpha).
\end{eqnarray}
The cloned state $W_{\rm clon}(\alpha)=W_{\rm clon}(x,p)$ is
a convolution of $W_{\rm in}$ with a bivariate Gaussian with the 
excess noise variances $\lambda_x$ and $\lambda_p$,
\begin{eqnarray}\label{Wcl}
W_{\rm clon}(x,p)&=&\frac{1}{2\pi\sqrt{\lambda_x\lambda_p}}
\int dx'dp' W_{\rm in}(x',p')\nonumber\\
&\times&\exp\left[-\frac{(x-x')^2}{2\lambda_x}
-\frac{(p-p')^2}{2\lambda_p}\right] \;,
\end{eqnarray}
where our choice $s=0$ leads in Eq.~(\ref{clone}) to excess noise
symmetric in phase space \cite{note}, 
$\lambda_x=\lambda_p=(M-1)/2M$.
Note that in our scales, a quadrature's vacuum variance is
$\case{1}{4}$, i.e., $[\hat{x},\hat{p}]=\case{i}{2}$.
Let us express the fidelity of the cloning process in the Wigner
representation,
\begin{eqnarray}\label{fidelity}
F\equiv\langle\psi_{\rm in}|\hat{\rho}_{\rm clon}|\psi_{\rm in}\rangle=
\pi\,\int d^2\alpha W_{\rm in}(\alpha)W_{\rm clon}(\alpha) \;.
\end{eqnarray}
Now we consider the Wigner function of a possibly squeezed Gaussian input state
(with mean values $x_0$ and $p_0$ and squeezing parameter $s$),
\begin{eqnarray}\label{Win}
W_{\rm in}(x,p)=\frac{2}{\pi}\exp[-2e^{-2s}(x-x_0)^2-2e^{2s}(p-p_0)^2].
\end{eqnarray}
Since the mean values are conserved through cloning, the fidelity
does not depend on $x_0$ and $p_0$, and  
without loss of generality we can set $x_0=p_0=0$.
Our MQC with $s=0$ exactly realizes optimum symmetric $1\rightarrow M$
telecloning of coherent states [$s=0$ also in Eq.~(\ref{Win})],
$F=F_{{\rm clon},1,M}^{{\rm coh\,st},\infty}=M/(2M-1)$ \cite{Cerf2}. 
Furthermore, the above protocol demonstrates that our MQC is
capable of transferring {\it arbitrary} quantum states $W_{\rm in}$
simultaneously to $M$ remote and separated receivers with equal minimum
excess noise in each output mode. 
Less excess noise emerging at each output for arbitary
$W_{\rm in}$ would imply that we could also beat the optimum-cloning
limit for coherent-state inputs.
Minimum excess noise symmetrically added in phase space
does not necessarily ensure
optimum telecloning {\it fidelities} at the outputs. It does for
coherent-state inputs, but squeezed-state inputs 
($s\neq 0$) require asymmetric excess noise, 
$\lambda_x=e^{2s}(M-1)/2M$,
$\lambda_p=e^{-2s}(M-1)/2M$,
according to $F=2/[\sqrt{(4\lambda_x e^{-4s}+2e^{-2s})
(4\lambda_p e^{4s}+2e^{2s})}]$ from Eq.~(\ref{fidelity}).
Adjusting parameter $s$ in $W_{\rm MQC}$ correspondingly
however, ensures optimum fidelities, just as for the local
``non-universal'' Gaussian cloner which has a similar $s$-dependence
\cite{local}.
The structure of $W_{\rm MQC}$ becomes clearer, when we look 
at the generation of this state.  

Using an ideal phase-free beam splitter operation $B_{12}(\theta)$
acting on two modes $\hat{c}_1$ and $\hat{c}_2$ as
$\hat c_1 \rightarrow \hat c_1 \sin\theta + \hat c_2\cos\theta$,
$\hat c_2 \rightarrow \hat c_1 \cos\theta - \hat c_2\sin\theta$,
we can define a sequence of beam splitters acting on $M$
modes (``$M$-splitter'' \cite{PvL1}) as 
$B_{M-1\,M}\left(\sin^{-1}1/\sqrt{2}\right)B_{M-2\,M-1}
\left(\sin^{-1}1/\sqrt{3}\right)\times\cdots\times
B_{12}\left(\sin^{-1}1/\sqrt{M}\right)$. 
The recipe to build an MQC is now as follows: 
first produce a bipartite entangled state by combining two squeezed vacua
(one squeezed in $p$ with $r_1$ and the other one squeezed in $x$ 
with $r_2$) at a phase-free beam splitter with 
reflectivity/transmittance parameter $\theta=\theta_0$.
Then keep one half as a ``port'' mode
(our mode 1) and send the other half together with $M-1$ ancilla
modes through an M-splitter.
The ancilla modes, 
$\hat{a}_i'=\cosh s\,\hat{a}_i+\sinh s\,\hat{a}_i^{\dagger}$ with
$\hat{a}_2$, $\hat{a}_3$, ..., $\hat{a}_M$ being vacuum modes,
are either vacua $s=0$ or squeezed vacua $s\neq 0$.
In the latter case, in order to obtain $W_{\rm MQC}$, also the squeezing 
of the two inputs of the first beam splitter needs to be modified
by the same amount (given by $s$).
The $s$ squeezers play exactly the same role as for local cloning
\cite{local}, namely to switch between different squeezing of
the input states.
These instructions imply that, although $W_{\rm MQC}$ is an entangled
multi-mode or multi-party state, it is actually {\it bipartite}
entanglement between mode 1 and the $M$ other modes that makes telecloning
possible.
The squeezing responsible for the entanglement corresponds to
$|10\log_{10}[(\sqrt{M}-1)/(\sqrt{M}+1)]|$ dB (if $r_1=r_2$), 
which is about 7.7 dB
for $M=2$, 5.7 dB for $M=3$, 4.8 dB for $M=4$, and 4.2 dB for $M=5$.
That the squeezing and hence the entanglement approaches zero as $M$ 
increases is consistent with the convergence of the optimum cloning 
fidelity $F_{{\rm clon},1,M}^{{\rm coh\,st},\infty}=M/(2M-1)$ 
to $F_{\rm class}=1/2$.

Their bipartite character is what $W_{\rm MQC}$ and the qubit
telecloning state proposed by Murao {\it et al.} \cite{Murao} have 
in common.
However, as opposed to $W_{\rm MQC}$ (except for $M=1$), 
the qubit state contains maximum bipartite entanglement.
On the other hand, the qubit states are in some sense more symmetric
and even more ``multiuser-friendly'', as they are actually
$2M$-partite states containing bipartite entanglement between
$M$ parties ``on the left side'' and $M$ parties ``on the right side''.
Due to this symmetry, {\it each} particle on {\it each} side can function 
as a ``port'' enabling the transfer of quantum information to all particles
on the other side \cite{Murao}. We can also construct such an MQC
for continuous variables with exactly the same
properties as the qubit state, but the price we have to pay is
that we need infinite squeezing, i.e., maximum bipartite entanglement
for any $M$. The corresponding $2M$-mode state is generated by
first producing an infinite-squeezing EPR state \cite{Sam1} and then
sending {\it both} halves each together with $M-1$ ancilla
modes through an M-splitter. 
Also this MQC enables optimum $1\rightarrow M$ telecloning
of coherent states, but instead of a fixed ``port'' mode, any mode
``on the left side'' built from the left EPR-half or
``on the right side'' built from the right EPR-half
can now function as a ``port'' for sending quantum information to the
other side. Let us emphasize the analogy between this particular
continuous-variable MQC and Murao {\it et al.'s} qubit telecloning state  
\cite{Murao} by displaying the former for $M=2$ and $s=0$
in the Schr\"{o}dinger representation:
\begin{eqnarray}\label{Muraolike}
|\psi_{MQC'}\rangle&\propto&\int dx\, dy\, dz\, \exp(-y^2-z^2)\nonumber\\
&\times& |x+y\rangle|x-y\rangle|x+z\rangle|x-z\rangle \;.
\end{eqnarray}
Despite its nice symmetry properties,
it is an unphysical (unnormalizable) state as opposed to the
state $W_{\rm MQC}$ which does without
infinite squeezing. Our results suggest that also for qubits,
a less symmetric but more economical version of an MQC might exist, 
since $F_{{\rm clon},1,M}^{{\rm univ},2}=(2M+1)/(3M)$ approaches 
$F_{\rm class}=2/3$ as $M$ increases. 

An important question now is if $W_{\rm MQC}$
is indeed the {\it most} economical version of an MQC. 
Does this state exploit minimal squeezing resources? 
At least the linear optics part, one beam splitter followed by
an $M$-splitter, is certainly the simplest possible choice.
Nevertheless, let us consider a much broader class of $M+1$-mode
states, namely all multipartite entangled states that can be generated
via quadratic interaction Hamiltonians (i.e., an arbitrary combination
of multi-port interferometers, squeezers, down-converters, etc.).
This arbitrary combination may be decomposed by Bloch-Messiah reduction 
\cite{Sam4} into a set of $M+1$ squeezers 
$\hat{a}_i'=\cosh \xi_i\,\hat{a}_i+\sinh \xi_i\,\hat{a}_i^{\dagger}$
with vacuum inputs $\hat{a}_i$,
and a subsequent linear multi-port [unitary transformation $U(M+1)$],
$\vec{b}=U(M+1)\vec{a'}$ with 
$\vec{b}=(\hat{b}_1,\hat{b}_2,...,\hat{b}_{M+1})^T$ etc.
Without loss of generality, mode $\hat{b}_1$ can be chosen as a ``port'',
and rather than assuming a phase-free symmetric beam splitter before
the ``Bell detection'', 
we consider now any unitary matrix $U(2)$
acting on the input mode $\hat{a}_{\rm in}$ and $\hat{b}_1$,
$(\hat{b}_{\rm u},\hat{b}_{\rm v})^T=U(2)
(\hat{a}_{\rm in},\hat{b}_{\rm 1})^T$.
An arbitrary unitary
matrix acting on $M+1$ modes can be decomposed into beam
splitters and phase shifters as \cite{Reck}
\begin{eqnarray}\label{generalU}
U(M+1)&=&(B_{M\,M+1}B_{M-1\,M+1}\cdots B_{1\,M+1}\nonumber\\
&&\times B_{M-1\,M}B_{M-2\,M}\cdots B_{12}D)^{-1} \;.
\end{eqnarray}
The $M(M+1)/2$ beam splitter operations each depend 
on two parameters, $B_{kl}(\theta_{kl},\phi_{kl})$, which is
here an $M+1$-dimensional identity matrix with the entries
$I_{kk}$, $I_{kl}$, $I_{lk}$, and $I_{ll}$ replaced by 
$e^{i\phi_{kl}}\sin\theta_{kl}$,
$e^{i\phi_{kl}}\cos\theta_{kl}$, $\cos\theta_{kl}$,
and $-\sin\theta_{kl}$, respectively.
All extra phase shifts have 
been put in matrix $D$ having diagonal elements 
$\beta_1,\beta_2,...,\beta_{M+1}$ and off-diagonal terms zero. 
The entire telecloning process based on this generalization
depends on $M^2+3M+6$ parameters. With an optimization algorithm based 
on a genetic code \cite{Dobbs}, we numerically confirmed for $M=2$ 
(16 parameters) and $M=3$ (24 parameters) that our MQC given by 
$W_{\rm MQC}$ uses the least total squeezing. 
In every calculation, the optimization 
(with excess noise symmetric in phase space being optimal) forces 
$M-1$ auxiliary modes to approach vacuum and only a pair of modes
to be squeezed, each mode by at least  
$10\log_{10}[(\sqrt{M}-1)/(\sqrt{M}+1)]$ dB (if equally squeezed,
otherwise less squeezing in one mode is at the expense of more squeezing
in the other mode, exactly as for our proposed state).

This work was funded in part under project QUICOV under the
IST-FET-QJPC programme and by a DAAD Doktorandenstipendium (P.v.L.).

\end{multicols}

\begin{references}
\bibitem{note} It is not easy to prove the validity of Eq.~(\ref{clone})
for arbitrary $M$, but one can verify it for small $M$. For
arbitrary $M$, one can rely on a proof in the Heisenberg picture.
\end{references}
\end{document}